\documentstyle[12pt]{article}

\newcommand{\be}{\begin{equation}}
\newcommand{\ee}{\end{equation}}
\newcommand{\bea}{\begin{eqnarray}}
\newcommand{\eea}{\end{eqnarray}}
\newcommand{\gsim}{\lower.7ex\hbox{$\;\stackrel{\textstyle>}{\sim}\;$}}
\newcommand{\lsim}{\lower.7ex\hbox{$\;\stackrel{\textstyle<}{\sim}\;$}}
\newcommand{\tev}{\,{\rm TeV}}
\newcommand{\gev}{\,{\rm GeV}}
\newcommand{\mev}{\,{\rm MeV}}

\newcommand{\ev}{\,{\rm eV}}
\newcommand{\mpl}{M_{pl}}
\newcommand{\mst}{M_{*}}
\newcommand{\mcr}{M_{cr}}

\begin{document}

\baselineskip=17pt
\pagestyle{plain}
\setcounter{page}{1}

\begin{titlepage}

\begin{flushright}
CLNS-98/1577 \\
SU-ITP-98/53\\
CERN-TH/98-267
\end{flushright}
\vspace{0mm}

\begin{center}
{\huge Black Holes and}
\vskip 2mm
{\huge Sub-millimeter Dimensions} 
\vspace{2mm}
\end{center}
\begin{center}
{\large Philip C. Argyres$^a$, Savas Dimopoulos$^b$,}\\
\vspace{2mm}
{\large and John March-Russell$^c$\footnote{On leave
of absence from the Institute of Advanced Study, Princeton, NJ}}\\
\vspace{2mm}
{\em $^a$ Newman Laboratory, Cornell University, Ithaca NY 14853,  
USA}\\
{\em $^b$ Physics Department, Stanford University, Stanford CA 94305,  
USA}\\
{\em $^c$ Theory Division, CERN, CH-1211, Geneva 23, Switzerland}
\end{center}
\vspace{0mm}
\begin{center}
{\large Abstract}
\end{center}
\noindent
Recently, a new framework for solving the hierarchy problem was
proposed which does not rely on low energy supersymmetry or
technicolor.  The fundamental Planck mass is at a TeV and the observed
weakness of gravity at long distances is due the existence of new
sub-millimeter spatial dimensions.  In this letter, we study how the
properties of black holes are altered in these theories.  Small black
holes---with Schwarzschild radii smaller than the size of the new
spatial dimensions---are quite different.  They are bigger, colder,
and longer-lived than a usual $(3+1)$-dimensional black hole of the
same mass.  Furthermore, they primarily decay into harmless bulk
graviton modes rather than standard-model degrees of freedom. 
We discuss the interplay of our scenario with the holographic
principle.  Our results also have implications for the bounds on
the spectrum of primordial black holes (PBHs) derived from the
photo-dissociation of primordial nucleosynthesis
products, distortion of the diffuse gamma-ray spectrum, overclosure of
the universe, gravitational lensing, as well as the phenomenology of
black hole production.  For example, the bound on the spectral index
of the primordial spectrum of density perturbations is relaxed from
1.25 to 1.45-1.60 depending on the epoch of the PBH formation.  In
these scenarios PBHs provide interesting dark matter candidates; for
6 extra dimensions MACHO candidates with mass $\sim 0.1M_\odot$ can
arise.  For 2 or 3 extra dimensions PBHs with mass $\sim 2000 M_\odot$
can occur and may act as both dark matter and seeds for early galaxy
and QSO formation.
\end{titlepage}
\newpage

\section{Introduction}

Recently, Arkani-Hamed {\it et al.}\ \cite{ADD,AADD,ADDlong} proposed
a new framework for solving the hierarchy problem which does not rely
on supersymmetry or technicolor.  The hierarchy problem is solved by
bringing the fundamental Planck scale, where gravity becomes
comparable in strength to the other interactions, down to the weak
scale. The observed weakness of gravity at long distances is due to
the presence of $n$ new spatial dimensions large compared to the
electroweak scale.  This follows from Gauss' Law which relates the
Planck scales of the $(4+n)$-dimensional theory $\mst$ and the
long-distance 4-dimensional theory $\mpl$,
\be
\mpl^2 \sim R^n \mst^{n+2},
\label{radrel}
\ee
where $R$ is the size of the extra dimensions.  Putting $\mst \sim 1$
TeV then yields
\be
R \sim 10^{{30\over n} - 17} \mbox{cm} .
\ee
For $n=1$, $R \sim 10^{13}$ cm, so this case is excluded since it
would modify Newtonian gravity at solar-system distances. Already for
$n=2$, however, $R \sim 1$ mm, which is precisely the distance where
present experimental measurements of gravitational strength forces
stop. As $n$ increases, $R$ approaches $(\tev)^{-1}$ distances,
albeit slowly: the case $n=6$ gives $R \sim (10\mev)^{-1}$.

While the gravitational force has not been measured beneath a
millimeter, the success of the SM up to $\sim 100\gev$ implies that
the SM fields can not feel these extra large dimensions; that is, they
must be stuck on a wall, or ``3-brane", in the higher dimensional
space. Summarizing, in this framework the universe is
$(4+n)$-dimensional with Planck scale near the weak scale, with $n
\geq 2$ new sub-millimeter sized dimensions where gravity and perhaps
other fields can freely propagate, but where the SM particles are
localised on a 3-brane in the higher-dimensional space.

An important question is the mechanism by which the SM fields are
localised to the brane. The most attractive possibility is to embed in
type I or type II string theory using the D-branes that naturally
occur \cite{AADD,Dbrane}. This has the obvious advantage of being
formulated within a consistent theory of gravity, with the additional
benefit that the localization of gauge theories on a 3-brane is
automatic \cite{Dbrane}.   Of course,
the most pressing issue is to ensure that this framework is not
experimentally excluded. This was the subject of \cite{ADDlong} where
phenomenological, astrophysical and cosmological constraints were
studied and found not to exclude the framework.

There are a number of important papers with related ideas which
concern themselves with the construction of string models with extra
dimensions larger than the string scale \cite{antoniadis,HW}, and with
gauge coupling unification in higher dimensions \cite{DDG,tye,bachas}.
There are also significant papers by Sundrum on the effective theory of
the low energy degrees of freedom in realizations of
the world-as-a-brane \cite{Raman}.

The objective of the present paper is to study one of the model
independent aspects of the new framework, namely that gravity is
altered at distances less than the size of the new dimensions. Since
these distances are always less than a millimeter, this change---as
explained in reference \cite{ADDlong}---is not important for normal
stars or for neutron stars. However, cosmologically interesting black
holes have routinely sub-millimeter sizes and therefore their
properties can be drastically altered if there are new sub-millimeter
dimensions.  For example, traditional 4-dimensional black holes
evaporating today weigh as much as Mount Everest and are a fermi
across. Their properties and signatures change radically in our
framework.  In section 2 we discuss the microphysical properties of
such ``small'' black holes, including the interplay of
the world-as-a-brane scenario with the holographic principle.
In the rest of the paper we discuss
possible observational implications for cosmology and astrophysics,
which include: implications for the bounds on the spectrum of
primordial black holes (PBHs) derived from the photo-dissociation of
BBN products, distortion of the diffuse gamma-ray spectrum,
overclosure of the universe, gravitational lensing, as well as the
phenomenology of black hole production.  For example, the bound on the
spectral index $N$ of the primordial spectrum of density perturbations is
relaxed from $N=1.25$ to $\sim 1.45-1.6$ depending on the epoch of the PBH
formation.  In these scenarios PBHs provide interesting dark matter
candidates; further, in the cases of $n=2$ or $3$ extra dimensions
PBHs with mass $\sim2000 M_\odot$ arise naturally
and may act as both dark matter and possibly even seeds for early 
galaxy and QSO formation, although we are not able to examine the
physics of this last suggestion in any detail.

\section{Properties of Small Black Holes}

When the Schwarzschild radius of a black hole is much smaller than the
radius $R$ of the compactified dimensions, it should be insensitive to
the brane and the boundary conditions in the $n$ transverse
dimensions, and so is well approximated by a $(4+n)$-dimensional
Schwarzschild black hole.  As we increase its size at some point its
radius exceeds $R$ and the black hole should go over to an effective
description as a 4-dimensional black hole.  Estimates made below will
use the value $\mpl=10^{19}\gev$ for the 4-dimensional Planck scale,
and for the $(4+n)$-dimensional Planck scale, $\mst$, values varying
between $1\tev$ for $3\le n \le 6$ and, for the $n=2$ case, the
astrophysically preferred value $\mst=10\tev$ (see \cite{ADDlong} for
discussion of the bounds on $\mst$).

We can understand the cross-over behavior more quantitatively by
recalling some basic properties of Schwarzschild black holes in $4+n$
dimensions.  We start with the size of small black holes.  Following
Laplace \cite{l1789} we can estimate the horizon radius $r_{s(4+n)}$
of a black hole of mass $M$ by equating the kinetic energy of a
particle moving at the speed of light with its gravitational binding
energy:
\be
{m c^2\over2} \sim {G M m \over r_{s(4+n)}^{1+n}}.
\ee
Setting $c$ and $\hbar$ to 1, and using the $(4+n)$-dimensional
relation between Newton's constant and the Planck scale,
\be
G = {1\over\mst^{2+n}} ,
\ee
gives
\be
r_{s(4+n)} \sim {1\over\mst} \left(M\over\mst\right)^{1\over n+1} .
\label{schrad}
\ee
Using the exact Schwarzschild solution in higher dimensional general
relativity \cite{highdBH}, modifies this relation only by numerical
factors:
\be
r_{s(4+n)} = {1\over\mst} \left(M\over\mst\right)^{1\over n+1} \cdot
\left(8 \Gamma((n+3)/2)\over (n+2) \pi^{(n+1)/2}\right)^{1/(n+1)} .
\label{exactR}
\ee

This is to be compared with the four-dimensional Schwarzschild
radius
\be
r_{s(4)} \sim {1\over\mpl}\left(M\over\mpl\right) \sim
{1\over\mst}\left(M\over\mst\right) {1\over(\mst R)^n},
\label{schrad4}
\ee
where we have used (\ref{radrel}).  Thus we have the relation
\be
\left({r_{s(4)}\over r_{s(4+n)}}\right) \sim
\left({r_{s(4+n)}\over R}\right)^n.
\label{radiusreln}
\ee
The derivation of this relation is valid when $r_{s(4+n)}\le R$.
We immediately learn that when $r_{s(4)} < R$ then
\be
r_{s(4)} < r_{s(n+4)} < R .
\ee
This confirms that the cross-over behavior between 4 and
$(4+n)$-dimensional black holes takes place smoothly, and implies that
small enough black holes of a given mass will be larger in a brane
universe than otherwise.  The mass $\mcr$ of a black hole right at
the cross-over region where $r_{s(4)} \sim r_{s(n+4)} \sim R$ is, from
(\ref{schrad}) and (\ref{radrel}),
\be
\mcr \sim \mpl \left(\mpl\over\mst\right)^{1+{2\over n}}
\sim 10^{{30\over n}-23} M_\odot
\label{mcross}
\ee
which ranges from about a hundredth of an earth mass for two extra
dimensions to that of a large building for large $n$.

Note that we have assumed that the brane tension itself does not
strongly perturb the $(4+n)$-dimensional black hole solutions that we
use by more than $O(1)$ factors.  We can argue this as follows
(specializing to the case of $n=2$ for simplicity---the details are
different for $n>2$, but the overall conclusions are unaltered): The
presence of the 3-brane in the 2-dimensional transverse space has the
effect of producing a conical singularity in the transverse space at
the site of the brane with deficit angle $\delta = f^4/4\mst^4$ (see
Ref.~\cite{Raman} for details).  The singularity itself is most likely
resolved by the $\tev$-scale string theory that underlies the
scenario, but most importantly at distances long compared to
$(1\tev)^{-1}$ the {\it only} effect of the brane is the deficit
angle.  Thus in the ``exact'' $(4+2)$-dimensional formulae the area of
the $S^{(n+2)}$ sphere $A(S^{(n+2)})$ should actually be replaced by
$(1-\delta/2\pi)A(S^{(n+2)})$.  But the deficit $\delta/2\pi$ must be
small for the consistency of the world-as-a-brane scenario.  The
general statement for any $n$ is that the curvature radius of the
internal dimensions must be larger than their extent, namely, $R$.
Moreover the dependence of the Schwarzschild radius of the
$(4+n)$-dimensional black hole on its mass $M$ and on $\mst$ is
unaltered by the presence of the brane.  We ignore these $O(1)$
correction factors in our discussion.

The Hawking temperature $T_{(4+n)}$ of a $(4+n)$-dimensional black
hole can be easily estimated from the first law of black hole
thermodynamics
\be
T_{(4+n)} = {dE\over dS} \sim {dM\over dA} \sim 
{M\over(r_{s(4+n)}\mst)^{n+2}}
\sim \mst\left(\mst\over M\right)^{1\over n+1} .
\label{thawk}
\ee
Using the exact Schwarzschild solution in higher dimensional general
relativity \cite{highdBH} again modifies this relation only by  
numerical factors:
\be
T_{(4+n)} = \mst\left(\mst\over M\right)^{1\over n+1} \cdot
\left((n+1)^{n+1}(n+2)\over 2^{2n+5} \pi^{(n+1)/2}
\Gamma((3+n)/2) \right)^{1/(n+1)} .
\label{exactT}
\ee
{}From (\ref{mcross}) the Hawking temperature for cross-over black  
holes goes as 
\be
T_{cr} \sim \mst (\mst/\mpl)^{2/n}.
\ee
Compared to the temperature $T_{(4)}\sim \mpl^2/M$ of a
4-dimensional black hole of the same mass, (\ref{thawk}) shows that
small black holes with mass $M<\mcr$ are cooler:
\be
T_{(4)} > T_{(4+n)} > T_{cr} .
\ee
This follows intuitively from the fact that they are larger, and
therefore have a larger entropy (area) for the same energy (mass).

The lifetime of a small black hole is correspondingly longer than that
of an equal mass 4-dimensional one.  The lifetime is estimated from
\be
{dE\over dt} \sim (\mbox{Area})\cdot T^{4+n}
\ee
giving from (\ref{schrad}) and (\ref{thawk})
\be
\tau_{(4+n)} \sim {1\over\mst} \left(M\over\mst\right)^{n+3\over n+1},
\ee
to be compared with the lifetime of a 4-dimensional black hole
\be
\tau_{(4)} \sim {1\over \mpl} \left(M\over\mpl\right)^3 .
\ee
Again, for $M<\mcr$ we find,
\be
\tau_{(4)} < \tau_{(4+n)} < \tau_{cr} .
\ee
Again, a more precise lifetime is derived using the higher
dimensional relationship between temperature and energy density,
\be
\rho(T) = g_* T^{n+4} \, {(n+3) \Gamma((n+4)/2) \zeta(n+4) \over
\pi^{(n+4)/2}},
\ee
where $\zeta$ is the standard Riemann zeta-function, and $g_*$ is the
number of effectively massless degrees of freedom in the bulk.  In the
minimal scenario in which only higher-dimensional gravity propagates
in the bulk, $g_*$ is just the number of polarization states of the
$(4+n)$-dimensional graviton
\be
g_* = {(n+4)(n+1)\over 2}.
\ee
The lifetime is then
\bea
\tau_{(n+4)} &=& {1\over\mst} \left(M\over\mst\right)^{n+3\over n+1}
\cdot \left(2^{2n^2+9n+13} \Gamma((n+3)/2)^{n+3} \over (n+2)^2 
\right)^{1/(n+1)} \cr
&&\qquad \cdot \left(\pi^{(2n+7)/2}\over (n+1)^{n+4}
(n+3)^2 \Gamma((n+6)/2) \zeta(n+4)\right) .
\label{exactime}
\eea

This last calculation implicitly assumed that the small black hole was
radiating in the bulk (off the brane) where it cannot emit any of the
particles localized to the brane.  If the black hole happened to
intersect the brane, its emission rate would be enhanced because of
the extra SM brane modes.  However this enhancement of radiation
channels on the brane as compared to the bulk gives a factor of at
most 20 and must be compared with the drastically reduced phase space
for radiating in the brane.  Since the width of the brane is less than
or on the order of $\mst^{-1}$, the phase space suppression factor is
less than or on the order of $(\mst R)^{-n} \sim (\mst/\mpl)^{-2} \sim
10^{-30}$ at cross-over.  We should note that this conclusion
depends on our assumption that the energy-density due to the brane
tension does not drastically distort the bulk black hole solution.

Also note that in the standard 4-dimensional picture the smallest
mass BH that one can discuss before having to worry about quantum
gravity, or string theory, is $M=\mpl=1.2\times10^{19}\gev$.  But
in our picture
one can describe black holes by semi-classical physics down to much
smaller masses of order the fundamental Planck scale $M=\mst=1 \tev$.

\subsection{State counting and implications for holography}

A fundamental principle of 4-dimensional BH physics asserts that there is
one quantum state per Planck area.  The natural generalization to
$(4+n)$-dimensions is that there is one quantum state per $(2+n)$-dimensional
Planck volume.  However, the fundamental Planck length is now $O(\tev^{-1})$,
which is 16 orders of magnitude bigger than the usual 4-dimensional
Planck length.  Therefore one might think that there are far
fewer degrees of freedom in our framework, and, for example
the holography principle \cite{tHBS} becomes much more restrictive.
This is not the case.  

To see this let us first consider a big black hole with Schwarzschild radius
larger than the size of the extra dimensions.  What is the entropy
of such a big BH calculated from the $(4+n)$-dimensional perspective?
Taking the transverse dimensions to be
an $n$-torus, the horizon of a large black hole
would be approximately an $S^2 \times T^n$ with the $S^2$ of area
$r_{s(4)}^2$ and the $T^n$ of volume $R^n$.  The entropy is
proportional to the total area of the horizon in $(4+n)$-dimensional
Planck units:
\be
S \sim A \sim r_{s(4)}^2 R^n \mst^{n+2}.
\ee
But using Eqn.~(\ref{radrel}) this is 
\be
S \sim  r_{s(4)}^2 \mpl^2,
\ee
exactly the entropy of the black hole as calculated from the
4-dimensional effective field theory perspective!  The reason for this
is that although the unit volume has vastly increased, we can
now fill up the volume of the extra dimensions with quantum states,
and the two effects precisely compensate each other as they should.
Similarly the temperatures of a large BH  as calculated from the two
perspectives agree
\be
T_{(4)} \sim
M/(r_{s(4)}^2 R^n \mst^{n+2}) \sim \mpl^2/ M.
\ee  

For small BH's of radius less than $R$, the 4- and $(4+n)$-dimensional
entropies of a BH of a given mass are related by
\be
S_{(4+n)} = S_{(4)}\left({R\over r_{s(4)}}\right)^{n/(n+1)}.
\label{entropyreln}
\ee
Since we have shown $r_{s(4)} < r_{s(4+n)} < R$, the $(4+n)$-dimensional
entropy is always larger than the 4-dimensional one.  Thus at distances
less than $R$ there are more quantum states available, rather than
fewer.  Finally, one may wonder what happens to the 4 vs. ${(4+n)}$-dimensional
pictures when $M\sim\mst$, and $r_{s(4+n)}\sim 1/\mst$.  At this mass
the $(4+n)$-dimensional BH has entropy $S_{(4+n)}=O(1)$, and roughly
1 quantum state.  This does not conflict with any 4-dimensional results
since for as small a mass as $1\tev$ the usual 4-dimensional BH
is very far below the mass where it can be described by anything other
than the full theory of quantum gravity (if the BH exists at all). 

In summary, for large BH's the holography constraints
coincide in 4 and $(4+n)$ dimensions, as they should, whereas
for small BH's the holography constraints are less restrictive
in $(4+n)$ dimensions despite the fact that the fundamental Planck
volume is now much bigger.

\section{Phenomenology of small black holes}

All black holes, no matter what their initial mass, in the end shrink
down to a size where they must be described by the $(4+n)$-dimensional
results already discussed.  Thus in the world-as-a-brane scenario
black holes are longer lived than they otherwise would be.  The most
important changes, however, involve the formation and decay of black
holes.  We now turn to an examination of how these modified properties
of small black holes in a brane universe affect various black hole
bounds on cosmological parameters.  These bounds can be divided into
three categories roughly concerned with (1) the decay of small black
holes, (2) the production of primordial black holes, and (3) the
present mass density of black holes.  We will address these in turn.

\subsection{Decay}

The bounds coming from the decay of black holes are enormously
weakened in a brane universe.  There are two factors which affect the
decay of black holes in a brane universe.  First, as we saw in the
last section, brane universe black holes are longer-lived than equal
mass black holes in a 4-dimensional universe would be.  For example,
from Eqn.~(\ref{exactime}), we find that the initial mass of black
holes evaporating today is $\simeq 5\times 10^{-27}M_\odot$ for $n=2$
extra dimensions, growing to $\simeq 2.4\times 10^{-19}M_\odot$ for
$n=6$.  These are to be compared to the initial 4-dimensional black
hole mass $\simeq 2.5\times 10^{-19}M_\odot$ that would be evaporating
today. 

The second, and much more important factor by far, is the relative
suppression of black hole radiation into the brane compared to
radiation into the bulk.  In greater detail, if the 3-brane thickness
relevant for standard model excitations is denoted by $t$ then, for
black holes smaller than the size $R$ of the extra dimensions, the
in-brane to bulk phase space suppression factor is roughly
$(t/r_{s(4+n)})^n$.  Certainly this is a very great suppression of the
visible modes until $r_{s(4+n)}$ approaches $t$.  At the very last
moments of the black hole's existence when $r_{s(4+n)}$ nears $t$ it
is possible that visible brane modes become unsuppressed.  Whether
this occurs depends on whether the black hole stays attached to the
brane when $r_{s(4+n)}<R$, or if it can wander off into the transverse
bulk dimensions.  Without knowing more details of the bulk and brane
theory it is not possible to calculate the probability of such
wandering in detail.  Nevertheless we would expect that as its horizon
size approaches $t$ random fluctuations in its radiation (with
momentum $\Delta p\sim t^{-1}$) would tend to make it leave the brane
in the late stages of its life.  Note, of course, that it is always
possible that the black hole formed in the bulk to start with.

Even in the most conservative case where the small black hole stays
localized on the 3-brane, the total amount of energy deposited over
the life of the black hole into SM modes is very greatly suppressed.
To see this note that the largest the 3-brane thickness $t$ can be
while still allowing a consistent phenomenology for the
world-as-a-brane scenario is $t=1\tev^{-1}$ (in principle much thinner
branes are also possible, which would give an even greater
suppression).  The usual constraints on the density of small black
holes following from Hawking evaporation involve the emission of
energetic standard-model particles such as photons which can disrupt
standard cosmology.  For instance photons more energetic that $\sim
1\mev$ can disassociate big bang nucleosynthesis products, ruining the
successful prediction of the light element abundances.  In the usual
4-dimensional case black holes have a Hawking temperature above
$1\mev$ for a mass of $M\simeq 10^{41}\gev$ or less.  Thus such black
holes can emit $O(10^{41}\gev)$ of energy into dangerous energetic
modes.  In contrast, even if the $(4+n)$-dimensional black hole stays
fixed to the brane it emits at most $O(1\tev)$ amount of energy into
standard model modes, a suppression factor of $10^{-38}$.  This means
that as far as a brane observer is concerned, black holes decay
essentially invisibly (only through gravitationally coupled modes),
with a possible $O(1\tev)$ flash of $\gamma$-rays at the last instant.
As a result, all of the evaporation constraints on the density of PBHs
are severely weakened in the world-as-a-brane scenario.  In summary,
small black hole decay is non-destructive, and there are no strong
limits from $\gamma$-rays, or BBN light element destruction.

The next subsection discusses the most significant remaining constraints
on PBH density---those that follow from overclosure.

\subsection{Production and Density Bounds}

The production of small black hole of a given mass in a brane
universe is easier than in a 4-dimensional universe.  This follows from
the fact that for $r_{s(4+n)} < R$, the $(4+n)$-dimensional
Schwarzschild radius is greater than their radius as calculated with
4-dimensional gravity, $r_{s(4+n)} > r_{s(4)}$, and thus a given mass
of matter on the brane has to be compressed less to form a horizon.
Such brane-localized matter has a pancake-like mass
distribution since its transverse extension is much smaller than its
in-brane extension.  Nevertheless this asymmetric collapse should lead
to a black hole because such a distribution is consistent with a
$(4+n)$-dimensional generalization of the hoop-conjecture for black
hole formation \cite{hoop} once the in-brane extent is less than
$r_{s(4+n)}$.

In principle there could be many astrophysical and cosmological
mechanisms that might form black holes.  However, given the model
dependence inherent in many of these mechanisms we will limit the
discussion to the relatively minimal possibility of production via
primordial density perturbations $\delta \equiv \delta\rho / \rho$,
possibly resulting from some period of inflation, although the origin
of the density perturbations will not affect our argument.

To start, let $T_{i}$ be the maximum temperature below which the
universe becomes the standard radiation dominated universe.  In
inflationary scenarios this is the reheating temperature after the end
of inflation.  The mass within the horizon at this epoch is given by
\be
M_{\rm H} \simeq 0.037 {\mpl^3\over g^{1/2}_* T_i^2}.
\label{horizonmass}
\ee
This is the smallest that the horizon mass can be in the era in which
the universe is normal.  According to standard arguments a black hole
forms when a density fluctuation $1/3 <\delta\rho/\rho <1$ enters the
horizon resulting in a black hole of the horizon mass
\cite{carr,primordBH}.  Our currently observable universe contains
many separate horizon regions from the time when $T=T_i$.  Thus an
average over the many horizon regions at $T=T_i$ determines the
properties of our universe.  As a result if the mean
$\delta\equiv\delta\rho/\rho$ at $T_i$ was $O(1)$ the entire universe
would form black holes massively overclosing the universe now.
Therefore the mean $\delta$ must be much less than unity, and PBHs are
only formed by rare fluctuations of $\delta$ away from its mean and
into the range $1/3 < \delta <1$.  To calculate the mass fraction of
the universe in black holes we must evaluate the fraction of early
horizons (at $T_i$) for which $\rho$ has fluctuated into the large
$\delta>1/3$ formation range.

The assumption of a Gaussian probability distribution $P(\delta)$ for
the density fluctuations $\delta(M)$ (where $\delta$ is expressed as a
function of the horizon mass $M$ at the wavelength of the
perturbation)
\be
P(\delta(M)) = { 1 \over \sqrt{2\pi\sigma_{rms}^2(M)}}
\exp \left(-{\delta^2(M)\over 2\sigma_{rms}^2(M)}\right),
\label{probdist}
\ee
enables a calculation of the initial mass fraction of black holes.
Here $\sigma_{rms}(M)$ is the mass variance evaluated at horizon
crossing.\footnote{Mild non-gaussianity---as for example studied in
the context of PBH formation in Ref.~\cite{primack}---does not
significantly alter the limits we find on the spectral index $N$. Also
$\sigma$ must be evaluated using a suitably defined window-function;
see \cite{carr,primordBH} for details.}

It is possible to correlate $\sigma_{rms}^2(M)$ on the small scales of
interest with the $\sigma_{rms}^2(M)$ measured at large scales by the
various cosmic microwave background experiments (COBE {\it etc.}).
This is done by making the standard assumption of a power law
behavior, of the form $k^{(N-1)}$ for the power spectrum, $N=1$
being the scale-invariant Harrison-Zeldovich spectrum.  Note that a
common use of the Hawking evaporation of small PBHs is to place limits
on the spectral index of models in which, as small scales are
approached, the size of the density perturbations increases.  Such
$N>1$ spectra are known as blue spectra and are not uncommon in the
attractive hybrid inflationary models.

As discussed in Refs.~\cite{carr,primordBH}, the COBE data constrains
the normalization of $\sigma_{rms}(M)$, leaving just the spectral
index $N$ of the density perturbations as a free parameter.  To a
sufficient approximation for our purposes, COBE implies the
normalization
\be
\sigma_{rms}(M) \simeq 10^{-4}\left( {M\over 10^{56}  
g}\right)^{(1-N)/4}.
\label{cobenorm}
\ee
The mass fraction in black holes, $\beta(M) \equiv \rho_{BH}(M) /
\rho_{tot}$, is just given by the probability that a fluctuation
reaches the formation range $1/3<\delta<1$.  From the Gaussian
probability distribution for $\delta(M)$, and the fact that for this
distribution PBH formation is exponentially closely concentrated near
the lower end of the formation range $\delta =1/3$, the initial mass
fraction is well approximated by
\be
\beta_i(M) = \int_{1/3}^1 P[\delta(M)] d\delta(M) \simeq  
\sigma_{rms}(M)\exp\left(-{1\over 18\sigma_{rms}^2(M)}\right).
\label{massfraci}
\ee 
Substituting Eqn.~(\ref{cobenorm}) then gives the predicted initial  
mass fraction as a function of the spectral index $N$.

Black holes formed by such density fluctuations are always very
massive compared to $T_i$, and so their density scales like
non-relativistic matter.  Taking account of the red-shifting of the
radiation bath, the black hole mass fraction today is given by
\be
\beta_0(M) = {T_i \over T_{eq}}\beta_i(M)
\label{massfraction}
\ee
where $T_{eq}\simeq 1\ev$ is the temperature of matter-radiation
equality, and we have used the fact that the BH's of interest are
cosmologically stable.  The statement
that mass $M$ black holes do not currently
overclose the universe is simply $\beta_0(M) < 1$.  The relation
Eq.~(\ref{horizonmass}) between the horizon mass at formation and the
temperature $T_i$, together with Eq.~(\ref{massfraction}) and the
expression for $\beta_i(M)$, Eq.~(\ref{massfraci}), then translates
into a limit on the spectral index $N$, given a value for $T_i$, the
maximum temperature at which the evolution of the universe is of the
normal radiation-dominated type.

With a blue spectrum of primordial density perturbations the
distribution function of PBHs is steeply concentrated at the smallest
possible scales, or equivalently the highest possible temperatures.
The lightest black holes that can be present with any significant
number density in our universe today are thus formed immediately after
the epoch of inflationary reheating.  In \cite{ADDlong,DMR} a
relatively conservative bound on the maximum possible reheat
temperature was derived by requiring that the gravitons radiated off
into the bulk by a (brane-localized) SM thermal bath not provide (an
effectively non-relativistic matter) bulk mass density that would
``overclose'' the universe.  As discussed in \cite{DMR}, this upper
bound on the temperature $T_i$ takes on the values $\sim(3, 5, 40,
170)\mev$ and $0.5\gev$ as the number of extra dimensions is varied
from $n=2$ to $n=6$.  (The analysis of \cite{DMR} shows that the more
stringent late photon decay constraint on the reheat temperature is
automatically avoided, allowing the temperatures $T_i$ quoted above.)
This translates to a horizon mass of $\sim 2\times 10^{60}\gev\sim
2000 M_\odot$ and $\sim 10^{56}\gev \sim 0.1 M_\odot$ for $n=2$ and
$n=6$, respectively.  Note that these masses, even for the $n=6$ case
are outside the range excluded by the MACHO and EROS microlensing
experiments---see for example Ref.~\cite{microlens}.  Indeed, the MACHO
and EROS collaborations report several events consistent with MACHO
masses in the range $M\ge 0.1M_\odot$.  Thus if the spectral index
is at the limit $N=1.47$ appropriate for $n=6$ extra dimensions (see
below), it is
possible that PBHs in these world-as-a-brane scenarios could be the
entire halo dark matter.  An interesting further possibility that is
worth mentioning is that for the lower $n$ cases where the horizon
mass is comparatively large, quasars, and maybe galaxy formation more
generally, may be seeded by these massive PBHs.  This is especially
interesting given the mounting evidence that substantial numbers of
QSOs form surprisingly early in the evolution of the universe.

In any case, from the constraint that PBHs do not overclose the
universe, together with the bounds on the maximum temperature $T_i$,
we can calculate the bounds on the amount of blue-shifting of the
spectral index $N$.  Applying the formulae of the previous paragraphs
leads to bounds lying between $N\lsim 1.59$ for 2 extra dimensions,
and $N\lsim 1.47$ for $n=6$.  These bounds on the amount of
blue-shifting are considerably {\it weaker} than those usually arising
from $(3+1)$-dimensional PBH production via primordial density
fluctuations.  If the MAP and PLANCK cosmic microwave background
satellite experiments measure a spectral index at the degree 
scale greater than $1.25$ then this would favor our framework.  Of course
if $N$ turns out to be less than the conventional bound $1.25$ then this
has no implications for the world-as-a-brane scenario, other than that the
number of primordial BH's produced from primordial density
fluctuations is very small.

\section*{Acknowledgements}
It is a pleasure to thank Nima Arkani-Hamed, Gia Dvali,
Alex Kusenko and Misha Shaposhnikov
for helpful comments and
discussions.  PCA and SD thank the CERN theory group for its
hospitality.  The work of PCA is supported in part by NSF grant
PHY-9513717 and an A.P. Sloan Foundation Fellowship.  The work of SD
is supported in part by NSF grant PHY-9219345-004.
The work of JMR is supported in part by
an A.P. Sloan Foundation Fellowship.

\def\pl#1#2#3{{\it Phys. Lett. }{\bf B#1~}(19#2)~#3}
\def\zp#1#2#3{{\it Z. Phys. }{\bf C#1~}(19#2)~#3}
\def\prl#1#2#3{{\it Phys. Rev. Lett. }{\bf #1~}(19#2)~#3}
\def\rmp#1#2#3{{\it Rev. Mod. Phys. }{\bf #1~}(19#2)~#3}
\def\prep#1#2#3{{\it Phys. Rep. }{\bf #1~}(19#2)~#3}
\def\pr#1#2#3{{\it Phys. Rev. }{\bf D#1~}(19#2)~#3}
\def\np#1#2#3{{\it Nucl. Phys. }{\bf B#1~}(19#2)~#3}
\def\mpl#1#2#3{{\it Mod. Phys. Lett. }{\bf #1~}(19#2)~#3}
\def\arnps#1#2#3{{\it Annu. Rev. Nucl. Part. Sci. }{\bf #1~}(19#2)~#3}
\def\sjnp#1#2#3{{\it Sov. J. Nucl. Phys. }{\bf #1~}(19#2)~#3}
\def\jetp#1#2#3{{\it JETP Lett. }{\bf #1~}(19#2)~#3}
\def\app#1#2#3{{\it Acta Phys. Polon. }{\bf #1~}(19#2)~#3}
\def\r nc#1#2#3{{\it Riv. Nuovo Cim. }{\bf #1~}(19#2)~#3}
\def\ap#1#2#3{{\it Ann. Phys. }{\bf #1~}(19#2)~#3}
\def\ptp#1#2#3{{\it Prog. Theor. Phys. }{\bf #1~}(19#2)~#3}

\end{document}